\documentclass[usenatbib]{basi}
\usepackage[T1]{fontenc}
\usepackage[british]{babel}
\usepackage[varg]{txfonts}

\def\beq{\begin{equation}}
\def\eeq{\end{equation}}
\def\bea{\begin{eqnarray}}
\def\eea{\end{eqnarray}}

\usepackage{rotating}
\usepackage{dcolumn}
\begin{document}
\title[Where will Einstein fail? Lessons for gravity and cosmology]
      {Where will Einstein fail? Lessons for gravity and cosmology}
\author[Niayesh Afshordi]{Niayesh Afshordi$^{1,2}$\thanks{email: \texttt{nafshordi@pitp.ca}
       \hspace{0.4cm}Article based on the Professor M.K. Vainu Bappu gold medal award (2008) lecture
                     given at IUCAA, Pune on 2011 October 15.}\\
       $^1$Department of Physics and Astronomy, University of Waterloo, 200 University Avenue West, Waterloo, \\
                     ON, N2L 3G1, Canada\\
       $^2$Perimeter Institute for Theoretical Physics, 31 Caroline St. N., Waterloo, ON, N2L 2Y5,Canada}

\setcounter{page}{1}
\pubyear{2012}
\volume{40}
\pagerange{\pageref{firstpage}--\pageref{lastpage}}

\date{Received 2012 March 14; accepted 2012 March 16}

\maketitle
\label{firstpage}

\begin{abstract}
 Einstein's theory of General Relativity is the benchmark example for empirical success and mathematical elegance in theoretical physics. 
 However, in spite of being the most successfully tested theory in physics, there are strong theoretical and observational arguments for why General Relativity should fail. It is not a question of if, but rather a question of where and when! I start by recounting the tremendous success in observational cosmology over the past three decades, that has led to the era of precision cosmology. I will then  summarize the pathologies in Einstein's theory of gravity, as the cornerstone of standard cosmological model. Attempts to address these pathologies are either inspired by mathematical elegance, or empirical falsifiability. Here, I provide different arguments for why a falsifiable solution should violate Lorentz symmetry, or revive ``gravitational aether''. Deviations from Einstein's gravity are then expected in: 1) cosmological matter-radiation transition, 2) neutron stars, 3) gravitomagnetic effect, 4) astrophysical black holes, and their potential connection to dark energy, and 5) early Universe, where the predictions are ranked by their degree of robustness and falsifiability. 
 
\end{abstract}


\section{Observational cosmology: the renaissance} 

Observational cosmology has come a long way since the early days of being ``the science of two numbers''. The
revolution could be traced back to the discovery of the cosmic microwave background (CMB) in 1964 by Penzias and Wilson \citep{1965ApJ...142..419P, 1965ApJ...142..414D}, which won them the Nobel prize in 1978, and confirmed an earlier prediction of George Gamow \citep{1948PhRv...73..803A}. Gamow had realized that the observed abundance of light nuclei can be explained from a hot uniform big bang phase, which is the inevitable beginning of an expanding universe in general relativity. However, it also predicts a relic background of radiation with $T \simeq 5$ K, which is reasonably close to the current observed value  of $T = 2.73$ K \citep[e.g.,][]{2009ApJ...707..916F}. 

Gamow's triumph could be arguably thought of as the beginning of physical cosmology, as it bore all the marks of scientific method as we know it: a self-consistent theory is devised based on observations (cosmic expansion and stellar elemental abundances), which makes predictions for completely different observables (microwave background radiation), that are later confirmed. 

The subsequent successes of observational cosmology only further confirmed the predictions of the big bang
paradigm, and its general relativistic framework. In particular, the observation of large scale correlations
in galaxy surveys, such as CfA, Las Campanas, 2dF, and the Sloan Digital Sky Survey confirmed the growth of
gravitational instability expected in a Friedmann-Robertson-Walker (FRW) space-time dominated by
non-relativistic matter. Observation of anisotropies in the CMB angular maps by the Cosmic Background
Explorer Differential Microwave Radiometer (COBE DMR) experiment connected the gravitational instability of structures today to their linear seeds in the early Universe \citep{1992ApJ...396L...1S}. A decade later, CMB observations reached their next milestone with the Wilkinson Microwave Anisotropy  Probe (WMAP), which measured the anisotropy power spectrum down to $0.1$ degrees with unprecedented precision \citep{2003ApJS..148..135H}, and verified consistency with the concordance cosmological model \citep{2003ApJS..148..175S}. Yet another leap in sensitivity is expected from Planck satellite in early 2013. 

In the mean time, geometrical tests in the late-time cosmology, in particular high redshift supernovae Ia as standard candles \citep{1998AJ....116.1009R,1999ApJ...517..565P}, and baryonic acoustic oscillations (BAO) in the spatial correlation of galaxies as standard rulers \citep{2005ApJ...633..560E}, confirmed a consistent picture for cosmic expansion history which appeared to have started accelerating some 5 billion years ago due to a mysterious dark energy component. 

These successes have led many to call the current era, the era of ``precision cosmology''. Indeed, a whole battery of cosmological observations, ranging from Lyman-$\alpha$ forest in quasar spectra on scales smaller than $1$ Mpc, to the statistics of galaxies and galaxy clusters on scales of $10-100$ Mpc, and CMB anisotropies on scales of $10-10^4$ Mpc are well-described by a six-parameter model \citep[e.g., see][]{2002PhRvD..66j3508T}, often dubbed as concordance or $\Lambda$CDM cosmological model (where $\Lambda$ and CDM stand for cosmological constant and cold dark matter, which comprise most of the energy in the present day Universe). These parameters include the present day densities of baryons, dark matter, and dark energy, as well as the amplitude and power of the initial spectrum of cosmological fluctuations (assuming that it is a power-law)\footnote{The sixth parameter is the optical depth for Thomson scattering to the last scattering surface of the CMB, which depends on the cosmic time when most atoms are reionized due to early star formation . While, in principle, this parameter is not independent of the others, it cannot be robustly calculated from our current models of star formation in the high redshift Universe.}. With the compilation of independent constraints from different cosmological observations, the statistical errors on these parameters have shrunk quickly over the past decade. Most of these parameters are now measured with 2-3\% statistical precision \citep[e.g.,][]{2011ApJS..192...18K}, even though it is sometimes hard to quantify the systematic errors on these measurements.     

The renaissance of observational cosmology has also been recognized by the greater physics community, who have awarded two Nobel prizes in Physics to this discipline in the past six years: The Nobel prize in 2006 was awarded to John Mather and George Smoot for discovery of blackbody spectrum and anisotropy of the CMB. The Nobel prize in 2011 was awarded to Saul Perlmutter, Adam Riess, and Brian Schmidt for the discovery of dark energy (or cosmological constant, the $\Lambda$ in $\Lambda$CDM) by study of magnitude-redshift relation of high redshift supernovae Ia. The two discoveries, however, were of very different nature: the first confirmed two of the most fundamental predictions of standard big bang model, while the latter revived an apparently redundant but enigmatic parameter in General Relativity.  As I will discuss below, this dichotomy is reminiscent of the status of modern cosmology today.   

\section{Is cosmology solved? The price of precision}

There is a famous quote from Lev Landau, the great Russian physicist of the 20th century, which says: ``Cosmologists are often in error, but seldom in doubt''! 

Unfortunately, Landau passed away in 1968, from the complications of a car accident that he was involved in 6 year earlier, so I wonder whether he ever had a chance to reflect on Penzias and Wilson's discovery of the CMB. Maybe, if he had, or if he had lived for another decade or so to witness the onset of the renaissance in observational cosmology, he might have revisited his original scepticism. Nevertheless, one often wonders whether there is some wisdom in the words of such great minds that would survive beyond the ages. For me, this rang particularly true in October 1998, when two of the most influential figures in modern cosmology, James Peebles and Michael Turner debated ``The Nature of the Universe'' in the main auditorium of Smithsonian's National Museum of Natural History in Washington, DC. The 1998 debate subtitle was ``Cosmology Solved?'', and their points of view appeared on the arXiv shortly afterwards \citep{1999PASP..111..274P,1999PASP..111..264T}. Despite being the early days of precision cosmology (which was a term also coined by Turner), and only a few months from the discovery of dark energy from supernovae Ia observations, Turner was very optimistic that the basic tenets of $\Lambda$CDM cosmology, along with inflation, will survive further scrutiny. On the other hand, Peebles was more cautious: ``We have a well defined, testable, and so
far quite successful theoretical description of the expansion: the relativistic Friedmann-Lemaitre cosmological 
model. The observational successes of this model are impressive
but I think hardly enough for a convincing scientific case.''. It appears that, as we discussed above, the influx of observational data over the ensuing decade has validated Turner's vision of precision cosmology. However, there is a price for this success which is often ignored.

More than 99 percent of today's energy content of the Universe in the concordance cosmological model is either unidentified, or invisible to us \citep[e.g., see][]{2004PhDT.........6A,2004ApJ...616..643F}! The most enigmatic component of $\Lambda$CDM is $\Lambda$  or the so-called dark energy, which comprises around 73\% of the energy of the Universe today. We will discuss the mysteries of $\Lambda$ (often called the cosmological constant problem) in the next section, as it is intimately connected to the quantum theories of high energy physics and even a quantum theory of gravity. 

The next biggest contributor is cold dark matter (or CDM) which makes up around 23\% of the cosmic energy budget. The most popular candidates for CDM are in the form of elementary particles: either weakly interacting massive particles (or WIMP's) or very light scalar particles, hypothesized to resolve the strong CP problem, known as axions. However, none of the efforts to find non-gravitational evidence for these particles have managed to conclusively detect these particles.  Therefore, it remains a possibility that  a more bizarre candidate, such as a modification of gravity, could explain the same observations.   While none of the proposed alternatives to CDM have enjoyed a similar phenomenological success in explaining both the early and late Universe observations \citep[e.g.,][]{2009CQGra..26n3001S}, apparent failures of CDM in matching observations on small scales may point to a more complex possibility \citep[e.g.,][]{2011MNRAS.415L..40B}.      

Even most of the standard model particles (often referred to as baryons), which comprise the remaining 5\% of the energy of the Universe, are expected to lie in a tenuous intergalactic medium, which has remained largely invisible to us. Attempts to account for these baryons in representative samples of the Universe, found in large galaxy clusters, has been controversial, and arguably misses up to 30\%-40\% of the cosmic baryonic budget \citep[e.g.,][]{2007MNRAS.378..293A,2011Sci...331.1576S}. 

Finally, inflation, a period of rapid exponential expansion in the very early Universe
\citep{1981PhRvD..23..347G,1982PhLB..108..389L}, which is often credited for generating a nearly spatially
flat cosmology with a nearly scale-invariant spectrum of cosmological fluctuations, is plagued by questions
of empirical falsifiability. It turns out that while natural expectations from (certain) inflationary models
are consistent with these observations, it is very easy to introduce arbitrarily large modifications to
inflationary predictions by modifying inflationary initial conditions, and/or adding extra  physics to the
inflationary model. In the absence of a(n established) UV-complete model of high energy physics (which
includes quantum gravity), it is hard to argue whether such modifications might (or might not) be natural
\citep[e.g.,][]{2006hep.th...12129C}. Further complication is introduced by the possibility of eternal
inflation, where predictions are plagued by the infamous ``measure problem'', which we will come back to in
Section \ref{falsify}.  

As a careful reader might have noticed, we have mentioned quantum gravity more than once in our introduction. This is not a coincidence, as we will see in the next section.

\section{(Cosmologist's) quantum gravity problems}\label{qg}

The search for a consistent quantum theory that includes gravity (or geometrodynamics) as a component is as old as both general relativity (which made geometry and gravity synonymous) and quantum mechanics \citep[see][for an overview of the history of quantum gravity]{2000gr.qc.....6061R}. By now, it has become quite clear that a quantization of Einstein's theory of relativity, while well-behaved as an effective field theory, is non-renormalizable and thus fails (at least as a perturbative theory) as gravitons approach Planck  energy ($ M_p c^2\equiv (\hbar c^5 / G_N )^{1/2} \simeq 1.22 \times 10^{19}$ GeV). It is easy to see this on dimensional grounds: Considering small perturbations around Minkowski background, $g_{\mu\nu}= \eta_{\mu\nu}+h_{\mu\nu}$, in natural units ($\hbar=c=1$), the GR action can be written as:
\beq
S_{GR} \sim -\frac{M^2_p}{16\pi}\int d^4x~ (h+\alpha_2 h^2+\alpha_3 h^3 + ...) \Box h,\label{s_gr}
\eeq
where $\alpha_n$'s represent dimensionless constants, we have abbreviated the tensorial structure of the
equation, and ignored additional subtleties in dealing with gauge symmetries. Considering the zero point
fluctuations of $h_{\mu\nu}$ on energy scale $E$, from the free or quadratic action (first term in equation \ref{s_gr}) we have:
\beq
\langle h^2 \rangle_E \simeq 8\pi M^{-2}_p \int^{\omega^{-1}(E)} \frac{d^3k}{(2\pi)^3 \omega(k)} \sim \left (E\over M_p\right)^2, \label{h2_lorentz}
\eeq
 where, in the last step, we have used the Lorentz-invariant (or Minkowski space) dispersion relation for
gravitons: $\omega(k) = k$. It is now quite clear that as $E$ approaches $M_p$, the perturbative expansion in
eauation (\ref{s_gr}) breaks down.

In quite the same way that $W^{\pm}$ and $Z$ gauge bosons  in modern electroweak theory cured the non-renormalizability of Fermi's low energy 4-point weak interaction,  various attempts at a theory of quantum gravity have mainly comprised of coming up with ``more fundamental'' degrees of freedom, that can be included in a renormalizable or finite theory. Such degrees of freedom could be fundamental strings \citep[e.g.,][]{1998stth.book.....P}, spin networks \citep{1995NuPhB.442..593R}, discrete causal sets \citep[e.g.,][]{1997IJTP...36.2759S}, or other discrete theories in space or space-time that resemble general relativity in  the continuum limit. Alternatively, it has been proposed that GR might be ``asymptotically safe'', i.e. it has a non-perturbative but well-defined quantization at arbitrarily high energies \citep{1979grec.conf..790W}. 

However, for most experimental physicists, approaching energies comparable to Planck energy \footnote{Here, we talk about energy per degree of freedom, or per particle. The total energy of macroscopic objects can obviously be far greater than Planck energy.}  is little more than a distant fantasy. The most powerful accelerators on Earth miss Planck energy by 15 orders of magnitude, while ultra high energy cosmic rays are still 9 orders of magnitude short of $M_p$. Therefore, the majority of physicists may not be much disturbed by the limitations of Einstein's theory of gravity.  

Unfortunately, astrophysicists do not enjoy such luxury! As first proved by Hawking and Penrose in a series of singularity theorems \citep{1970RSPSA.314..529H}, general relativity predicts its own demise! In particular, the end state of massive stars are in singularities, where temperatures exceed Planck energy. Millions of these singularities just happen to live in our own Galaxy, although they are expected to be shrouded by the event horizons of astrophysical black holes. While strong theoretical motivations exist for a cosmic censorship conjecture, which hides singularities behind the event horizons, there is no guaranty that this will be the case in a theory of quantum gravity. In fact, it is widely believed that event horizons do not exist in a full theory of quantum gravity, although the minimal quantum effects \citep[such as Hawking radiation,][]{1975CMaPh..43..199H} are far from being observable for astrophysical black holes. 

More seriously, as we approach big bang in our past, the temperature rises to Planck energy and beyond. Therefore, any consistent theory of cosmological initial conditions has to include quantum gravity, which impacts scalar and tensor gravitational degrees of freedom directly probed in observational cosmology (and potentially gravitational wave detectors)\footnote{It turns out that alternatives to the standard big bang, such as inflation or ekpyrotic scenarios, even though  may not approach Planck temperature, still rely on (speculative) non-perturbative features of a quantum theory of gravity.}.   

Yet, the most dramatic challenge of quantum gravity for cosmology does NOT come from Planck-scale physics. Quite to the contrary, it comes on scales that should see very little quantum corrections to general relativity. Like quantum gravity itself, this issue also dates back to the early days of the development of quantum mechanics. As early as 1920's, Pauli had recognized the tremendous amount of energy in the zero-point fluctuations of the electromagnetic field \citep{2002gr.qc.....8027S}. While this energy density is divergent, apparently, he had regulated the divergence by taking the classical radius of electron as a momentum cut-off. If all this energy were to gravitate according to Einstein's theory of relativity, the Universe would curve so much that it could not fit the lunar orbit, let alone the solar system, or the rest of the Galaxy! This is now recognized as the ``old cosmological constant (CC) problem''.

A more careful computation of vacuum energy involves introducing a Lorentz-invariant regulator, which suggests that a natural value for vacuum energy is roughly given by the sum of the fourth power of the particle masses in the theory:
\beq
 \rho_{\rm vac} \sim \sum_i \pm m^4_i.
\eeq 
For standard model, the sum is clearly dominated by the most massive particle, i.e. top quark with $m_t = 171$ GeV. If the energy $\rho_{\rm vac} \sim m^4_t$ were to gravitate, the entire Universe would have been smaller than a centimetre in size! Classical contributions from the Higgs potential can change this expectation by order unity, but short of a conspiracy, it is extremely unnatural to expect a cancelation between different contributions to $\rho_{\rm vac}$ to 1 part in $10^{60}$, in order to be consistent with the observed size of the Universe ($\sim 10$ Gpc). 

Because of the clear pathology in the above estimates, physicists ignored the old CC problem for the better part of the 20th century. More conscious theorists speculated a yet unknown symmetry that would lead to cancellations of different contributions to the vacuum energy. While examples of such symmetries, such as conformal symmetry or supersymmetry, exist, they all seem to be violated by the mass terms in standard model, and thus, as we argued, stop short of curing the old CC problem by some 60 orders of magnitude. In an influential review article \citep{1989RvMP...61....1W}, Steven Weinberg reviewed different approaches to the CC problem, outlining why each approach fails to solve the problem, while stopping short of dismissing any approach completely. He further speculated that, if all other approaches fail, anthropic considerations, which require a $\Lambda$ suitable for existence of intelligent life, will predict a value just within reach of cosmological observations. As we discussed in the previous sections, this was indeed verified at the turn of the century with the discovery of cosmic acceleration from high redshift supernovae.  

The modern CC problem is sometimes divided into three problems:
\begin{enumerate}
\item {\bf The Old CC problem:} Why isn't $\Lambda$ as big as its natural scale in particle physics?
\item {\bf The New CC problem:} Why does the observed vacuum energy have such an un-naturally small but non-vanishing value?
\item {\bf The Coincidence problem:} Why do we happen to observe vacuum density to be so close to matter density, even though their ratio can vary by up to 120 orders of magnitude during the cosmic history?
\end{enumerate} 
Let us reiterate that these are still problems (or puzzles) in quantum gravity, as they appear when we couple gravity to quantum mechanical theories, even though they concern physics far below the Planck scale. 

Other than the {\it gravitational aether} model that we will discuss in Section \ref{aether}, the only known ``solution'' to the CC problem comes from the anthropic considerations, or its modern incarnations in the string landscape and/or eternal inflation. In fact, as we mentioned above, one may argue that the non-vanishing $\Lambda$ was ``predicted''  based on these considerations. In the next section, we discuss to what extent this can be compared to predictions in other  scientific theories.  

\section{Anthropic landscape: physics vs falsifiability}\label{falsify}

Presumably the best (and the worst!) way to define physics is by the set of problems that are tackled by practicing physicists. Nevertheless, a careful study of the history of meaningful progress in science, as often done by philosophers, may reveal common features that may not be immediately obvious to practicing scientists. Probably the most influential philosopher of science of the 20th century was Karl Popper, who coined the term {\it critical rationalism} to describe his philosophy of a scientific theory \citep{1992sbwl.book.....P}. Popper argues that scientific theories are in fact measured by their falsifiability, as opposed to their verification, as no amount of positive tests can in fact verify a theory. However, a single counter-example suffices to rule out (or falsify) a theory.  More significantly, a scientific theory needs to be falsifiable, i.e. there should exist observations and/or experiments that can potentially falsify the theory. In practice, falsifiable theories that survive more tests than their competitors will be considered more successful.     

So, do anthropic considerations provide a falsifiable solution to the CC problem? To answer this, it is interesting to recount how developments in string theory and cosmology culminated in the anthropic principle at the turn of the century.

One of the most puzzling features of the standard big bang scenario was how points in the Universe that have never been in casual contact have the same temperature to 1 part in $10^5$. This is known as the ``horizon problem'', which was later solved by cosmic inflation, as we discussed in the previous section \citep{1981PhRvD..23..347G,1982PhLB..108..389L}. What inflation does is to stretch a small causally connected patch exponentially, across many Hubble horizons. So, points that appear causally disconnected today were parts of the same Hubble patch at the beginning of inflation. However, it was later realized that for many (if not most) successful inflationary models, inflation never ends! While the success of standard big bang theory requires inflation to end before big bang nucleosynthesis, there are always regions in field space that never stop inflating in these model. Since the physical volume of these regions is exponentially bigger than those that have stopped inflating, most of the volume of the Universe will always be inflating. This is known as eternal inflation, and has a significant anti-Copernican implication: If our cosmology emerges from eternal inflation, we cannot live in a typical region of the Universe \citep[e.g.,][]{1994PhRvD..49.1783L}.   

The second development was when research in string theory, which is the most popular contender for a theory of quantum gravity, failed to single out any particular vacuum for the theory. In fact, it was argued that string theory might have as many as $10^{500}$ viable vacua (or a landscape of vacua), with each vacuum having a very different low energy physics \citep[e.g.,][]{2004JHEP...01..060A}. 

The combination is now straightforward: No matter where you start the Universe in the string landscape,
assuming that it permits eternal inflation, there will be an infinite time to populate all the other
landscape vacua via quantum tunneling.  Of course, most of these regions will not be hospitable to humans (and presumably other intelligent life). Therefore, we can use anthropic principle\footnote{Weinberg calls this application the ``weak anthropic principle''.}, to pick the region where cosmology and particle physics allows humans to live. In particular, this region cannot have a very big positive or negative $\Lambda$, as neither allow enough time for galaxies, stars, planets, and life (as we know them) to form \citep{1989RvMP...61....1W}.     

Besides the nebulosity of the notion of ``intelligent life'', one of the problems with this interpretation is that all the predictions are now probabilistic. However, unlike quantum mechanics, where we compute probabilities for finite ensembles, the eternally inflating ensembles are inherently infinite in size. This is known as the ``measure problem'' (which we referred to earlier), as you could find very different probabilities, depending on how you regulate (or cut off) your infinite ensembles. 

The second problem is what we started this section with, i.e. falsifiability. Since most of the string landscape exists beyond our cosmological horizon, or at energies much higher than those accessible in accelerators, it is really hard to test (or falsify) its existence. Of course, we might get lucky and see glimpses of another vacuum \citep[e.g.,][]{2011PhRvL.107g1301F}, but for the most part, it has been hard to come up with ways to falsify this paradigm \citep[but see][for a notable possible exception]{2012arXiv1202.5037K}.  
 
We should emphasize that string theory and inflation are the direct results of extending {\it locality} and {\it unitarity}, the underlying principles of relativity and quantum mechanics, to gravity and cosmology. My personal point of view is that their failure in coming up with a falsifiable cosmological model (and most notably a falsifiable solution to the CC problem\footnote{We should note that none of the popular alternatives to string theory have been particularly more successful in addressing the CC problem.}), is cause to revisit these sacred principles of 20th century physics. We will discuss this next.
 
\section{Why aether? I. Quantum gravity and early Universe}\label{horava}

It has been long recognized that one of the ways to deal with infinities in quantum gravity (and even infinities in quantum field theory), is to break Lorentz symmetry \citep[see e.g.,][]{2006AnPhy.321..150J}. The reason is that Lorentz group SO(3,1), unlike e.g., the rotation group SO(3), is non-compact, and thus has an infinite volume. Therefore, the sum over the intermediate states for many quantum mechanical calculations (for rates or energy shifts) yield infinities. For renormalizable theories, these infinities can be absorbed in renormalization of a finite number of parameters that can be fixed empirically. However, this is not the case for non-renormalizable theories, such as gravity, which require renormalizing an infinite number of parameters, rendering the theory non-predictive. 

It is easy to see how violating Lorentz symmetry can cure this problem \citep{2009PhRvD..79h4008H}. Going
back to our quantized gravitons in Section \ref{qg}, we can see that using $\omega(k) = k^3/M^2$ in equation (\ref{h2_lorentz}) yields:
\beq
\langle h^2 \rangle_E \simeq 8\pi M^{-2}_p \int^{\omega^{-1}(E)} \frac{d^3k}{(2\pi)^3 \omega(k)} \sim \left(M\over M_p\right)^2 \ln\left(E\over M\right).
\eeq
Therefore, as long as $M \ll M_p$, the theory remains perturbative, even for energies far beyond $M_p$.  The
anisotropic scaling of space and time in this theory (as $\omega \propto k^3$) is known as Lifshitz
symmetry\footnote{While dispersion relation is enough to describe the quadratic action, 
\citet{2009PhRvD..79h4008H}  went on to write most general non-linear actions which obey (local or global)
Lifshitz symmetry and spatial diffeomorphism invariance. This is known as Ho{\v r}ava-Lifshitz gravity.}. A more realistic scenario is an interpolation between Lorentz symmetry at low energies, and Lifshitz symmetry at high energies:
\beq
\omega(k) = k + \frac{k^3}{M^2}.\label{omega_k}
\eeq 

Even though Lorentz symmetry is approximately recovered in the IR, there is still a single  preferred frame
in which equation (\ref{omega_k}) can be valid, as the dispersion relation is not invariant under Lorentz
transformation. This amounts to a revival of ``gravitational aether'', as an additional component to the
geometric structure of Einstein's gravity. Moreover, equation (\ref{omega_k}) violates the {\it locality} of relativistic theories, as localized perturbations can travel arbitrarily fast.  

Note that this already resolves the ``horizon problem'', which was one of the original motivations for
cosmological inflation. Moreover, the dispersion relation $\omega(k) = k^3/M^2$ leads to a scale-invariant
spectrum of cosmological fluctuations, provided that it can be converted to curvature perturbations at late
times \citep{2009JCAP...06..001M}. In other words, with Lifshitz symmetry we can potentially kill two birds
with one stone: make gravity renormalizable and generate the correct statistical distribution for
cosmological fluctuations. However, more detailed studies are necessary to quantify the phenomenological
implications of Ho{\v r}ava-Lifshitz cosmology as an alternative to cosmic inflation.  

In the next section, we provide a {\it second} motivation for aether, based on a falsifiable solution to the CC problem.   
  
\section{Why aether? II. Cosmological constant problem} \label{aether}

The old CC problem can be quantified as the pathologically large contribution to $\rho_{\rm vac}$ in the stress tensor $T_{\mu\nu}$ on the right hand side of Einstein equations:
\beq
T_{\mu\nu} = \rho_{\rm vac} g_{\mu\nu} + ~{\rm excitations}.
\eeq
We thus see that if only the traceless part of $T_{\mu\nu}$ appeared on the right hand side of Einstein equations, gravity would not be sensitive to $\rho_{\rm vac}$, which could potentially resolve (at least) the (old) CC problem.  Let us write this as \citep{2008arXiv0807.2639A}:
\beq
(8\pi G') G_{\mu\nu} = T_{\mu\nu}-\frac{1}{4}g_{\mu\nu}T^\alpha_\alpha+T'_{\mu\nu}. \label{grav_aether}
\eeq
The reason we added $T'_{\mu\nu}$ to the right hand side of equation (\ref{grav_aether}) is that, thanks to
Bianchi identities and energy-momentum conservation, Einstein and stress tensors have zero divergence.
However, $T^\alpha_\alpha$ is not generically a constant, which implies that consistency of equation (\ref{grav_aether}) requires:
\beq
\nabla^{\mu}T'_{\mu\nu} = \frac{1}{4}\nabla_\nu T^\mu_\mu. \label{aether_cons}
\eeq
Here, we call $T'_{\mu\nu}$ ``gravitational aether'', which can be interpreted as an additional fluid (or degree of freedom) of this new theory of gravity.  

Of course, the predictions of the theory depend our choice for $T'_{\mu\nu}$. Given that equation
(\ref{aether_cons}) provides 4 equations, they completely specify the evolution of a perfect fluid with a
given equation of state. If we want to avoid introducing additional dimensionful scales to the theory, the
aether equation of state will be set by a constant: $w'=p'/\rho'$. It turns out that only two possibilities
are consistent with conditions of stability and element abundances from big bang nucleosynthesis: $w'=-1$ or
$w'>5$  \citep{2008arXiv0807.2639A}. The former possibility leads to the so-called unimodular gravity, which
has been discussed by several authors, including Einstein himself \citep[e.g., see][]{1989RvMP...61....1W}.
In this case, aether does not have any dynamical degree of freedom. However, equation (\ref{aether_cons}) can be solved to show that the CC re-emerges in the theory as an integration constant.

The second possibility is more novel and interesting. One may dismiss $w'>5$ due to superluminal propagation, as sound speed is $c_s = w'^{1/2} >1$ for aether. While, as we argued in the previous section, this should not necessarily scare us, the case of $w'\rightarrow \infty$, or the incompressible limit is particularly interesting.  It can be argued that, in this limit, sound waves in aether cannot be excited, and thus there is no superluminal signal propagation. One way to see this is that any phonon of finite wavelength has an energy  of $E=\hbar \omega = \hbar c_s k \rightarrow \infty$, which implies that we need infinite energy to excite aether phonons.  Furthermore, similar to the case of $w'=-1$, $w'=\infty$ does not have any {\it independent} dynamical degree of freedom\footnote{This statement is only strictly valid for an irrotational aether.}, even though it does have a velocity, and thus specifies a preferred frame at each point in space (Relativistic dynamics of irrotational incompressible fluids, otherwise known as cuscuton, has been studied in \citet{2007PhRvD..75h3513A}, and \citet{2007PhRvD..75l3509A}).     
    
Notice that the ``gravitational aether'' theory, as we just specified with $w'=\infty$, i.e.
\beq
(8\pi G') G_{\mu\nu} = T_{\mu\nu}-\frac{1}{4}g_{\mu\nu}T^\alpha_\alpha+ p'(u'_\mu u'_\nu -g_{\mu\nu}),\label{full_aether}
\eeq
has no additional parameter (or independent dynamical degree of freedom), compared to Einstein's gravity. So could it be consistent with all the precision and cosmological tests of gravity?

\subsection{Cosmology}

\begin{figure}
\centering
\includegraphics[width=0.75\textwidth]{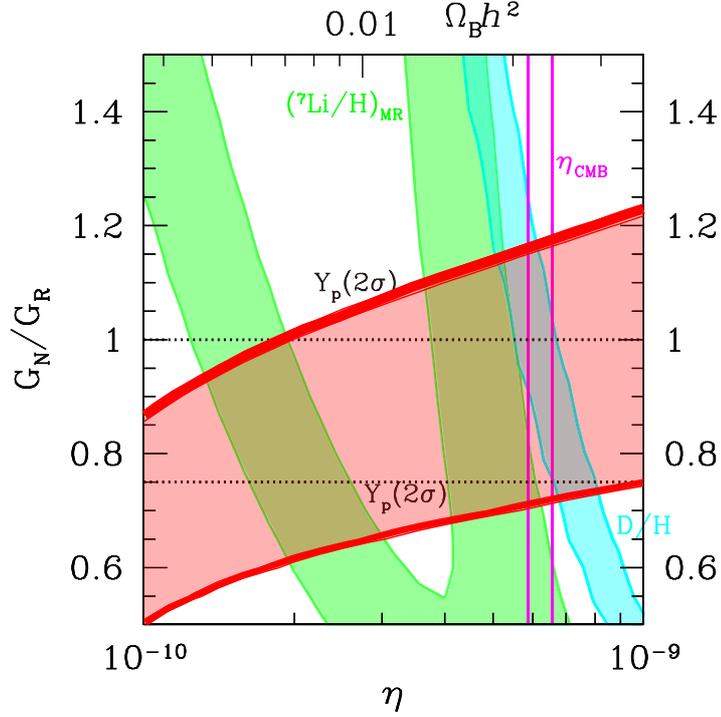}
\caption{Allowed regions with 2$\sigma$ lines for D/H, $Y_p$(or $^4$He) and
$^{7}$Li/H are shown \citep{2011PhRvD..84j3522A}. The upper and lower horizontal dashed
lines indicate GR and gravitational aether predictions,
respectively.}
\label{fig_BBN}
\end{figure}

Probably the sharpest prediction of the aether theory is that effective gravitational constant in Friedmann equation becomes dependent on the matter equation of state:
\beq
H^2=\frac{8\pi G_{\rm eff}}{3} \rho_m, G_{\rm eff} = (1+w_m) G_N,\label{cosmo}
\eeq
where $w_m=p_m/\rho_m$, is the matter equation of state, and $G_N$ is Newton's gravitational constant. In particular, this predicts that gravitational constant during radiation era was 33\% larger than in the matter era: $G_N/G_R= 3/4$. Fig. (\ref{fig_BBN}) and Table (\ref{tab-constraints}) summarize the big bang nucleosynthesis and CMB+late time cosmology constraints on this ratio \citep{2011PhRvD..84j3522A}.  We see that some datasets, namely $^7$Li, (most) CMB experiments, and Lyman-$\alpha$ forest in quasar spectra prefer ratios close to aether prediction, while others are closer to GR ($G_N=G_R$), or cannot distinguish between the two. Interestingly, however, all the best fits are at $G_N<G_R$\footnote{In the cosmological parameter estimation literature, this is often quantified as an observed effective number of neutrinos larger than 3, which is more than the number expected from the standard model of particle physics \citep[e.g.,][]{2011ApJS..192...18K}}. 

The influx of observational data, and in particular CMB power spectrum from the Planck satellite over the next year, should dramatically improve these constraints, and hopefully confirm or rule out aether predictions conclusively.   

\begin{table}
\caption{Summary of the constraints on $G_N/G_R$ and the associated $95\%$ confidence intervals for different
combinations of observational data \citep{2011PhRvD..84j3522A}. Here, WMAP, ACT, and SPT are different CMB
experiments, BAO stands for baryonic acoustic oscillations, while Sne and Hubble refer to measurements of
distance-redshift relation from supernovae Ia, and Hubble constant. Ly-$\alpha$ refers to measurements of
Lyman-$\alpha$ forest absorption distribution in the spectra of distant quasars. The last two rows take
Helium abundance $Y_p$ as a free parameter, while the rest fix it to its value set by Galactic observations.}
\centering
\begin{tabular}{l c }
\hline
&   ~~~$G_N / G_R$~~~ \\ \hline
WMAP+ACT & $0.73^{+0.31}_{-0.21}$ \\ 
WMAP+ACT+SPT & $ 0.88^{+0.17}_{-0.13}$  \\ 
WMAP+ACT+Hubble+BAO+Sne & $0.89^{+0.13}_{-0.11} $  \\ 
WMAP+ACT+SPT+Hubble+BAO+Sne & $0.94^{+0.10}_{-0.09}$  \\ 
WMAP+ACT+Sne+Ly-$\alpha$ (free $Y_p$) & $0.68^{+0.32}_{-0.25} $  \\ 
WMAP+ACT+SPT+Sne+Ly-$\alpha$ (free $Y_p$) & $0.90^{+0.27}_{-0.23} $  \\ \hline
\end{tabular}

\label{tab-constraints}
\end{table}

\subsection{Precision tests of gravity}

It can be shown that the simple scaling of effective Newton's constant with matter equation of state
(equation \ref{cosmo}) is still valid for inhomogenous situations, provided that:
\beq
u'_\mu=u_\mu, w_{m} = {\rm const.} \Rightarrow G_{\rm eff} = (1+w_m) G_N,
\eeq 
i.e. if the matter equation of state is constant, all the solutions in GR also satisfy the equations in the gravitation aether theory with a renormalized gravitational constant, if aether moves with matter. 

Let us first ignore the gravitational effect of local vorticity in matter and aether flows. In this regime, the flow of aether is completely fixed by matter, and we find that the only effect of aether is to renormalize Newton's constant by a factor of $1+w_m$. \citet{2011PhRvD..84j3522A} show that  none of the current precision tests of gravity constrain this effect, as it involves probing the internal structure of objects with near-relativistic pressure. We will discuss the case of Neutron stars in the next section. 

Coming back to the effect of vorticity, it turns out that rotational motion of aether is essentially decoupled from matter. Therefore, there is no reason for aether to rotate within rotating bodies. Assuming an irrotational aether will then boost the gravitomagnetic effect sourced by local vorticity, by 33\%, which is currently consistent at $2\sigma$ with LAGEOS and GPB experiments \citep{2011PhRvD..84j3522A}.  

\subsection{Neutron stars}

\begin{figure}[t]
\centering
   \includegraphics[width=0.75\linewidth]{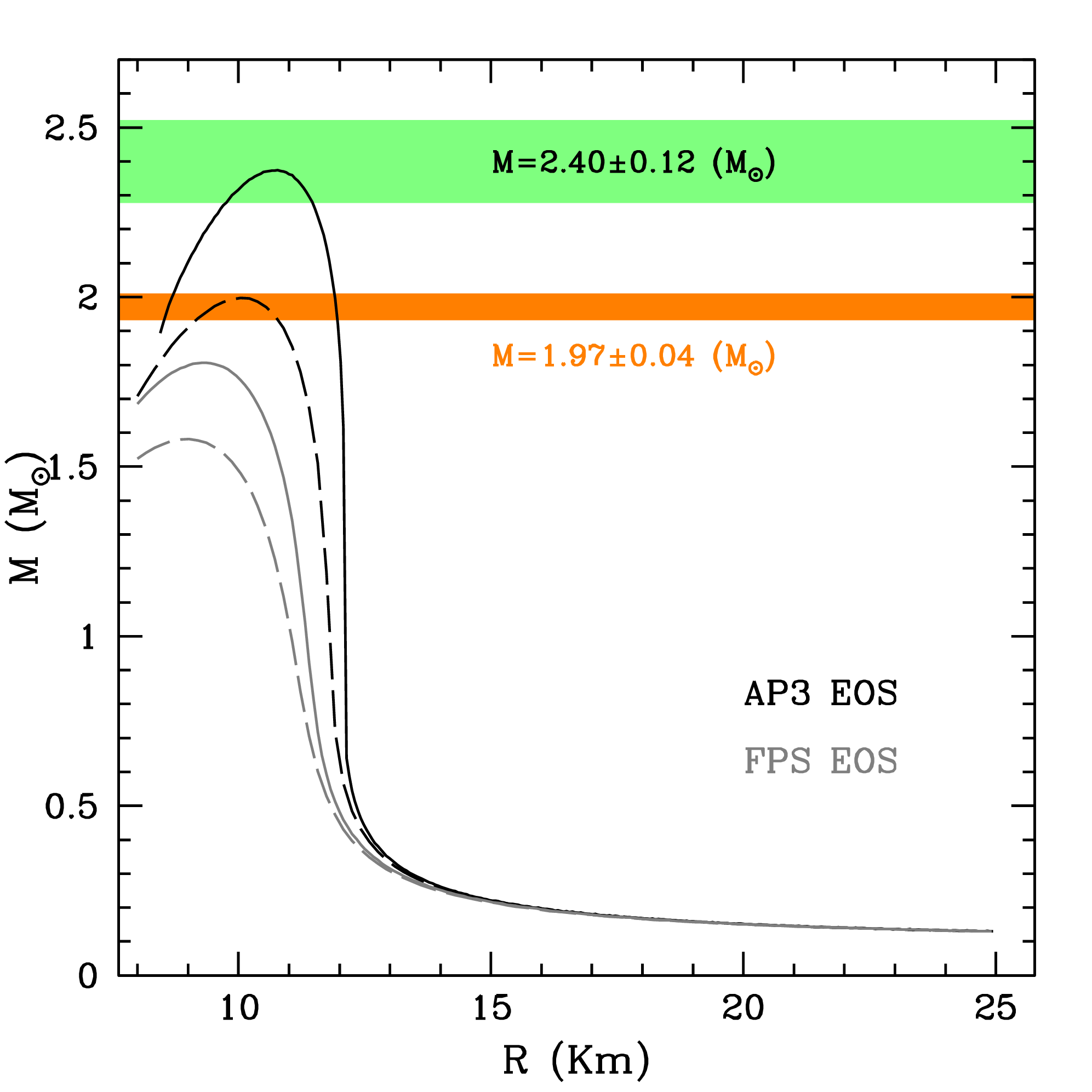}
\caption{The mass-radius relation of neutron stars given by general relativity (solid) and the aether theory (dashed) based on the parametrized AP3 (black) and FPS (grey) equations of state \citep{2011PhRvD..84f3011K}. The two observed pulsar masses of \citet{2010Natur.467.1081D} and \citet{2011ApJ...728...95V} (which has significantly more uncertainty) are shown in orange and green respectively.}
\label{fig:MREOS}
\end{figure}

As we mentioned above, the internal structure of objects with relativistic pressure is expected to be significantly different in the aether theory. The only known (close to equilibrium) astrophysical objects  that have this property are neutron stars. In Fig. (\ref{fig:MREOS}), we show the mass-radius relation for GR and aether theories, for two widely used nuclear equations of state \citep{2011PhRvD..84f3011K}. Most notably, we see that as gravity gets stronger with relativistic pressure, the maximum allowed mass of neutron stars (the so-called Oppenheimer-Volkov limit) decreases in aether theory. This is already close to ruling out the theory, for the most massive neutron star with a reliable mass measurement: $1.97 \pm 0.04 M_{\odot}$ \citep{2010Natur.467.1081D}.  However, the uncertainty in the nuclear equations of states may prohibit drawing definite conclusions from such observations. 

\subsection{Black holes: an explanation for dark energy?}

Probably the most speculative, and yet most fascinating, feature of the aether theory is how it couples to astrophysical black holes. As we discussed above, the exciting physics of singularities of black holes is hidden behind their event horizons in general relativity. In fact, astrophysical black holes are expected to have only two numbers (or hairs) associated with them: mass and angular momentum. However, this does not need to be the case for a different theory of gravity, such as the aether theory. 

By solving the static spherically symmetric aether equations in vacuum, \citet{2009PhRvD..80d3513P}  show
that the aether pressure, no matter how small at infinity, blows up close to the horizon of black holes. This
is not too surprising; one expects the same thing to happen for other types of matter. Of course, the reason
this is pathological for regular matter is that we do not expect matter to sit at rest close to the horizon,
but rather it would fall through. The story, however, is different for an incompressible fluid, as fluid
inside the light horizon can communicate with the outside\footnote{Communication is not meant here in its
literal sense, since incompressible fluids don't propagate signals. Nevertheless, the build-up of pressure
inside the horizon can impact the fluid equations outside.}. One can show that aether pressure would still
blow up even in a dynamical situation, just inside the horizon in a collapsing star (Saravani, Afshordi 
\& Mann, in preparation). What \citet{2009PhRvD..80d3513P}  propose instead is that this singularity is regulated by quantum gravity effects. 

\begin{figure}
\centering
\includegraphics[width=0.75\linewidth]{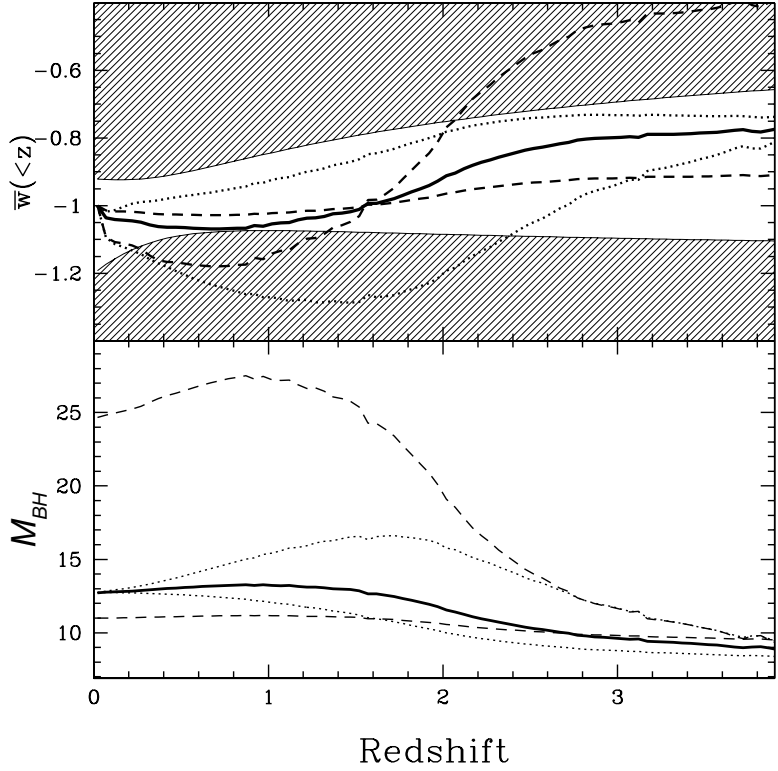}
 \caption
{From \citet{2009PhRvD..80d3513P}:
{\bf Bottom panel:} The mass-weighted geometric mean of black hole
masses, $M_{BH}$, in units of $M_{\odot}$ as a function of redshift.  Different lines represent different astrophysical scenarios of black hole formation.  {\bf Top panel:} The prediction of
these scenarios for the effective dark energy equation of state $\bar{w}(<z)$, given that aether pressure scales as $M_{BH}^{-3}$, which can be compared to constraints from cosmology.  The shaded area shows the region currently excluded at 68\% confidence level for this
parameter, as measured from cosmological observations \citep{2009ApJS..180..330K}.
}\label{fig-obsmstar}
\end{figure}

The vacuum metric in the presence of aether is given by \citep{2009PhRvD..80d3513P} :
\beq
ds^2= (1-r_s/r)\left[1+4\pi G_N p_0 f(r)\right]^2-(1-r_s/r)^{-1} dr^2-r^2d\Omega^2,\label{BH_metric}
\eeq
where $r_s=2G_NM_{BH}$ is the Schwarzschild radius, and $p_0$ is the aether pressure far away from the black holes. While $f(r)$ is an analytic function with a closed form, it is particularly illuminating to consider it close to $r_s$:
\beq
f(r) = r_s^2\left\{-2(r/r_s-1)^{-1/2}+{\cal O}\left[(r/r_s-1)^{1/2}\right]\right\}.
\eeq
We notice that unlike Schwarzschild black hole, the gravitational redshift $=g_{00}^{-1/2}$ approaches a maximum value at $r=r_s$: 
\beq
1+z_{\rm max} = -(8\pi G_N p_0 r_s^2)^{-1},
\eeq
while the metric is not defined for $r<r_s$\footnote{One can see the singularity just beyond $r=r_s$ by analytically continuing  metric (\ref{BH_metric}) in terms of proper radial distance.}. If we assume quantum gravity effects set a maximum gravitational redshift of Planck energy divided by Hawking temperature (which is equivalent to assuming that aether only becomes important within a Planck length of the horizon), the aether pressure away from the black hole is fixed to:
\beq
p_0 = - \frac{M^7_p}{256 \pi^2 M_{BH}^3} = p_{\rm obs, \Lambda} \left(M_{BH} \over 85 M_{\odot}\right)^{-3},\label{aether_DE}
\eeq  
i.e. for stellar black holes, which happen to make (by number or mass) the majority of astrophysical black holes in our Universe, this criterion fixes the pressure of aether to be comparable to the observed pressure of dark energy\footnote{It is not important to match the dark energy scale exactly with this prescription, as the exact definition of Planck scale can vary by an order of magnitude.}. Given that the average mass of astrophysical black holes evolves with redshift, one can predict an evolution for dark energy, or the dark energy equation of state (see Fig. \ref{fig-obsmstar}). 

However, extending the analysis of \citet{2009PhRvD..80d3513P} to more realistic situations, i.e. including
multiple moving black holes, in the presence of matter, has proven incredibly challenging. This is not too
surprising, as one needs to solve time-dependent non-linear partial differential equations on vastly
disparate scales (ranging from Schwarzschild to Hubble radii), and is practically impossible, in lieu of (a
yet to be found) appropriate approximation scheme. Until that is done, which can then provide falsifiable
predictions for the black hole-dark energy connection in the theory, equation (\ref{aether_DE}) remains little more than a(n extremely suggestive) numerical coincidence. 
   
\section{Lessons, challenges and outlook}

In this article, I have provided a very subjective overview of successes and failures of the standard cosmological model, especially in its relation to the fundamental physics of gravity. While observational cosmology is undergoing its renaissance, deep mysteries continue to haunt our theoretical understanding of the ingredients of the concordance cosmological model. In my opinion, the most enigmatic of these puzzles is the cosmological constant (CC) problem: the apparently extremely fine-tuned quantum vacuum energy, its relation to the observed dark energy, and possibly a more fundamental theory of quantum gravity.   

Fuelled by parallel theoretical invention of eternal inflation and string landscape at the end of the last
century, the anthropic ``solution'' to these puzzles is gaining more traction among the theoretical physics
community. Personally, I find this to be an alarming development. This is partly due to (near) lack of
falsifiability in these paradigms, which is a minimum standard for our scientific theories, as I discussed in
Section \ref{falsify}. Yet another source of worry is extrapolating well-established physical principles far beyond the regime in which they are tested, which happens both in eternal inflation and string theory. However, my most serious concern is that a general acceptance of an unfalsifiable framework will inevitably stifle a search for alternative solutions to these fundamental puzzles. If this happens, and observational probes of dark energy fail to find any deviation from a cosmological constant, we might very well enter a long period of intellectual stagnation in our understanding of the Universe, with no clear end in sight \citep[to borrow terminology from Christopher Stubbs,][]{2009qcfp.book.....T}.    

Given that a century of scientific enquiry to solve problems of quantum gravity, particularly the CC problem, have failed to come up with a falsifiable solution, it stands to reason that it might be time to revisit some of the fundamental principles of 20th century physics, i.e. locality (or Lorentz symmetry), and/or unitarity. In this article, I gave two arguments, based on renormalizability of gravity and the CC problem, for why we might have to give up Lorentz symmetry to come up with a falsifiable solution. I then outlined different predictions of the theory, with varying degrees of robustness, which can rule out or confirm the theory in comparison with GR, potentially as early as next year.  

Alternative approaches that take a similar point of view in regards to Lorentz symmetry include Einstein-Aether theories \citep{2004PhRvD..70b4003J}, Horava-Lifshitz gravity \citep{2009PhRvD..79h4008H}, and Shape Dynamics \citep{2011CQGra..28d5005G}. A typical objection to Lorentz-violating theories is that nothing prevents quantum corrections from introducing order unity violation of Lorentz symmetry at low energies \citep[e.g.,][]{2012PhRvD..85b4051L}. So why does particle physics seem to obey Lorentz symmetry to such precision? Here is an argument for why expectation of ${\cal O}(1)$ Lorentz violation might be too naive:

Imagine that aether is described by the gradient of a light scalar field $\chi$ which has a canonical kinetic term. If $m_{\chi} < H$, then the field will be slowly rolling down its potential, and thus $\nabla_\mu\chi$ specifies a preferred frame, which coincides with the cosmological comoving hypersurfaces. A typical Lagrangian which spontaneously breaks Lorentz symmetry for field $\phi$ can be written as:
\beq
{\cal L} = \frac{1}{2}(\partial \phi)^2-\frac{1}{2} m_\phi^2\phi^2+\frac{\left(\partial_\mu\phi\partial^\mu\chi\right)^2}{\Lambda^4}+ \frac{1}{2}(\partial \chi)^2-\frac{1}{2} m_\chi^2\chi^2. \label{lagrangian}
\eeq  
Given that the kinetic energy of the aether field $\chi$ should be less than the critical density of the Universe, we have:
\beq
\dot{\chi}^2 < M_p^2H^2,  
\eeq
which puts an upper limit on the Lorentz-violation for the $\phi$ field:
\beq
\delta c_{\phi} = \frac{\dot{\chi}^2}{\Lambda^4} < \left(M_p H \over \Lambda^2\right)^2,
\eeq  
Given that $\Lambda$ is the energy cut-off of the effective field theory in equation (\ref{lagrangian}), i.e. the theory is only valid for $E < \Lambda$, which puts an upper limit on the Lorentz-violation for $\phi$ particles at energy $E$, assuming the validity of the effective field theory:
\beq
   \delta c_{\phi} < \left(M_p H \over E^2\right)^2 \sim \left(E \over{\rm meV}\right)^{-4}.
\eeq   
We see that, already for energies as low as $\sim $ MeV (which is the mass of electron, and the typical energy of solar or reactor neutrinos), the Lorentz violation should be less than $10^{-36}$, and is further constrained for more energetic particles. This observation suggests that expectations of large violations of Lorentz symmetry might be too naive within a consistent effective field theory framework, which include gravity in a cosmological spacetime. 

Another objection, specific to the gravitational aether model (equation \ref{full_aether}) is that an action principle that could lead to these equations is so far non-existent. However, an action is only necessary if we want to quantize gravity, while the field equations (\ref{full_aether}), assuming that they can be consistently solved, are sufficient at the level of classical and semi-classical gravity. Presumably, more structure will be necessary to define a quantum theory that reduces to the gravitational aether  in the classical regime. 

Finally, I should mention previous attempts to decouple quantum vacuum from gravity, which have provided much of the inspiration for this work. These include massive gravity and degravitation \citep{2007PhRvD..76h4006D}, cascading gravity \citep[e.g.,][]{2008JCAP...02..011D}, and supersymmetric large extra dimensions \citep[e.g.,][]{2004AIPC..743..417B}. However, to the best of my knowledge, none of these frameworks have been developed well enough to make concrete cosmological predictions (at least in the regime that they address the CC problem). 

Let me close this article by stating the obvious, that Physics is an empirical science. 
While the bulk of activity in theoretical physics and astrophysics is driven by  (for lack of a better word) fashion, the credibility of a theory is ultimately judged by its concrete predictions against Nature, {\it not} its popularity, mathematical elegance, parsimony, etc. Do you hold your favourite theories to this standard?!

\section*{Acknowledgements}

I would like to first thank the Astronomical Society of India for awarding me its 2008 Vainu Bappu gold medal.  
Moreover, I am indebted to my students and collaborators Siavash Aslanbeigi,  Michael Balogh, Brendan Foster, Farbod Kamiab, Kazunory Kohri, Chanda Prescod-Weinstein, and Georg Robbers, who are responsible for most of the results that are reviewed in this article. I am supported by the University of Waterloo and the
Perimeter Institute for Theoretical Physics. Research at Perimeter
Institute is supported by the Government of Canada through Industry
Canada and by the Province of Ontario through the Ministry of Research
\& Innovation.


%

\label{lastpage}
\end{document}